\documentclass{article}
\usepackage{graphicx}
\usepackage{amsmath}

\def\ie{{\it i.e.},~}
\def\eg{{\it e.g.},~}
\def\etal{{\it et al.}~}
\def\4he{$^4$He}
\def\3he{$^3$He}
\def\7li{$^7$Li}
\def\Yp{Y$_{\rm P}$~}

\def\hii{H\thinspace{$\scriptstyle{\rm II}$}~}

\def\Nnu{N$_{\nu}$~}
\newcommand{\epm}{\ensuremath{e^{\pm}\;}}
\newcommand{\be}{\begin{equation}}
\newcommand{\ee}{\end{equation}}
\newcommand{\Deln}{\ensuremath{\Delta N_\nu\;}}
\def\Nnu{N$_{\nu}$~}

\newcommand\la{\lower0.6ex\vbox{\hbox{\ensuremath{\buildrel{\textstyle<}\over{\sim}\ }}}}
\newcommand\ga{\lower0.6ex\vbox{\hbox{\ensuremath{\buildrel{\textstyle>}\over{\sim}\ }}}}

\newcommand{\obh}{\ensuremath{\Omega_{\rm B} h^2\;}}

\begin{document}
 
\baselineskip 18pt

\begin{center}
{\Large {\bf  Neutrinos And Big Bang Nucleosynthesis }}
 \vskip5mm Gary Steigman\footnote{Email address: steigman@mps.ohio-state.edu} 
 \vskip3mm 
 \mbox{}%
 Departments of Physics and Astronomy, The Ohio State University,
Columbus, OH 43210, USA

\bigskip

\begin{abstract}

The early universe provides a unique laboratory for probing the frontiers
of particle physics in general and neutrino physics in particular.  The
primordial abundances of the relic nuclei produced during the first few
minutes of the evolution of the Universe depend on the electron neutrinos
through the charged-current weak interactions among neutrons and protons
(and electrons and positrons and neutrinos), and on all flavors of neutrinos 
through their contributions to the total energy density which regulates 
the universal expansion rate.  The latter contribution also plays a role 
in determining the spectrum of the temperature fluctuations imprinted 
on the Cosmic Background Radiation (CBR) some 400 thousand years later.  
Using deuterium as a baryometer and helium-4 as a chronometer, the 
predictions of BBN and the CBR are compared to observations.  The 
successes of, as well as challenges to the standard models of particle 
physics and cosmology are identified.  While systematic uncertainties 
may be the source of some of the current tensions, it could be that the 
data are pointing the way to new physics.  In particular, BBN and the
CBR are used to address the questions of whether or not the relic neutrinos
were fully populated in the early universe and, to limit the magnitude
of any lepton asymmetry which may be concealed in the neutrinos.

\bigskip


\end{abstract}
\end{center}
\newpage

\section*{1. Introduction}
During its early evolution the universe was hot and dense, passing
brief epochs as a universal particle accelerator and as a cosmic nuclear 
reactor.  As a consequence, through its evolution the entire universe 
provides a valuable alternative to terrestrial accelerators and reactors 
as probes of fundamental physics at the highest energies and densities.
Several decades of progress have validated this Particle Astrophysics
and Particle Cosmology approach to testing and constraining models
of High Energy Physics and Cosmology; for early work see, \eg 
\cite{wein,steig1,zeld}.  This strategy has proven especially
useful in connection with the physics of neutrinos (\eg masses,
mixing, number of flavors, etc.).

Neutrinos play two different, but equally important roles in Big Bang 
Nucleosynthesis (BBN).  On the one hand, electron-type neutrinos (and 
antineutrinos), through their charged current, weak interactions help
to regulate the neutron-proton ratio, which plays a key role in determining 
the abundance of helium (\4he) emerging from BBN when the universe is 
$\sim 20$~minutes old.  For example, since the \4he yield is largely
fixed by the supply of neutrons available at BBN, an asymmetry between 
$\nu_{e}$ and $\bar{\nu}_{e}$ (neutrino ``degeneracy'') will drive the 
relative abundance of neutrons up or down, thereby increasing or 
decreasing the relic abundance of \4he.  On the other hand, {\bf all} 
flavors of neutrinos were relativistic at BBN ($\sim$few MeV~$\ga~T~\ga 
~30$~keV), contributing significantly to the total density, which determines 
the expansion rate of the universe at that epoch.  The competition between 
the universal expansion rate (the Hubble parameter, $H$) and the nuclear 
and weak interaction rates is key to regulating the relic abundances 
of the light nuclides (D, \3he, \4he, \7li) synthesized during BBN.

This latter effect of (light, relativistic) neutrinos on the expansion 
rate also plays a role some 400 kyr later, at ``recombination'' (protons 
and electrons combine to form neutral hydrogen) when the Cosmic Background 
Radiation (CBR) is set free from the tyranny of electron scattering to 
propagate throughout the Universe.  By influencing the age of the Universe 
and the size of the sound horizon at recombination, the neutrinos help 
to fix the scales of the CBR temperature anisotropies observed by WMAP 
and other CBR detectors; see, \eg \cite{sperg} and references therein.
Here, however, neutrino degeneracy plays no role except, perhaps, by 
increasing the neutrino energy density and, thereby, affecting the 
expansion rate.  This latter effect is, generally, subdominant.

Since neutrinos influence the early evolution of the Universe at
these two, widely separated epochs ($\sim$~20 minutes and $\sim$
400 kyr later), the relics from BBN (light nuclides) and the
temperature anisotropies imprinted on the CBR provide two, largely 
independent windows on neutrino physics.  These connections and
what we have learned from them are reviewed here.  For further
details and references, see~\cite{barger1,barger2}; this review 
is largely based on these two papers.  After introducing some
notation in the next section, the constraints from the CBR are 
reviewed in \S3.  \S4 provides an overview of BBN and of the
current status of the comparisons between the observational data
and the predictions of the standard model (SBBN) as well as of
some general extensions of the the standard model (non-standard
BBN).  In \S5 the constraints from the CBR and from BBN are 
combined to identify the allowed ranges of the baryon and neutrino
parameters.  We conclude in \S6 with a summary and with an 
identification of the successes of the standard models of 
particle physics and cosmology and of some of the challenges 
confronting them. 

\section*{2. Notation}

To set the scene for the discussion to follow, it is useful to 
first introduce some notation.  We are interested in three, key 
parameters: the baryon density, the number of ``equivalent'' 
neutrinos, and a measure of a neutrino-antineutrino asymmetry.  

As the universe expands, the baryon density decreases.  A 
dimensionless measure of the baryon density is provided by 
the ratio of baryons to photons (in the CBR).  Following \epm 
annihilation, this ratio is preserved during the subsequent 
evolution of the universe.  The parameter $\eta$ is defined by 
the present (\ie post-BBN, post-recombination) value of this 
ratio: $\eta \equiv (n_{\rm B}/n_{\gamma})_{0}$; $\eta_{10} 
\equiv 10^{10}\eta$.  An equivalent measure of the baryon 
density is provided by the baryon density parameter, 
$\Omega_{\rm B}$, the ratio (at present) of the baryon density 
to the critical density.  In terms of the present value of the 
Hubble parameter, $H_{0} \equiv 100h$~km~s$^{-1}$~Mpc$^{-1}$, 
these two measures are related by 
\be
\eta_{10} \equiv 10^{10}(n_{\rm B}/n_{\gamma})_{0} = 
274\Omega_{\rm B}h^{2}.
\ee

In the standard model of particle physics there are three families of 
light neutrinos (\Nnu = 3) which, in the standard model of cosmology, 
are relativistic at BBN and also at recombination.  The early evolution
of the universe is ``radiation dominated'', \ie the energy density is 
dominated by the contributions from relativistic particles, including 
the neutrinos.  The universal expansion rate, as measured by the 
Hubble parameter, depends on the density: $H \propto \rho^{1/2}$.
Any additional (non-standard) contributions to the energy density
(such as, \eg from additional flavors of neutrinos) will result 
in a speed-up of the expansion rate,
\be
S \equiv H'/H =(\rho'/\rho)^{1/2} > 1.
\ee
{\it Any} non-standard contribution to the density may be written 
in terms of what would be the energy density due to an equivalent 
number of ``extra'' neutrinos $\Delta$N$_{\nu}$ (\Nnu $\equiv 3 + 
\Delta{\rm N}_{\nu}$).  Prior to \epm annihilation, this may be 
written as 
\be
\frac{\rho'}{\rho} \equiv 1 +  \frac{7\Delta{\rm N}_{\nu}}{43}.
\ee
Thus, either $S$, the expansion rate factor or, $\Delta{\rm N}_{\nu}$, 
the number of equivalent neutrinos, provide equally good measures 
of the early universe expansion rate.  While it is easy to imagine 
{\it extra} contributions to the energy density from new physics 
beyond the standard model, it must be noted that it is possible 
for \Deln to be negative, leading to a slower than standard, early 
universe expansion rate ($S < 1$).  For example, models where 
the decay of a massive particle, produced earlier in the evolution 
of the universe, reheats the universe to a temperature which is 
not high enough to (re)populate a thermal spectrum of the standard 
neutrinos ($T_{\rm RH}~\la 7$~MeV), will result in \Deln $< 0$ and 
$S < 1$ \cite{lowreheat}.

For any neutrino flavor {\it i}, an asymmetry (``neutrino degeneracy'')
between the numbers of $\nu_{i}$ and $\bar\nu_{i}$, relative to the 
number of CBR photons, can be quantified by the net lepton number 
$L_{i}$, the neutrino chemical potential $\mu_{i}$ or, by the 
dimensionless degeneracy parameter $\xi_{i} \equiv \mu_{i}/T$:
\be
L_{i} \equiv \frac{n_{\nu_{i}}-n_{\bar\nu_{i}}}{n_\gamma}= 
\frac{\pi^2}{12 \zeta(3)}\bigg(\xi_{i}+\frac{\xi_{i}^3}{\pi^2}\bigg)\,.
\ee
Although we are interested in lepton asymmetries which are orders of 
magnitude larger than the baryon asymmetry (B~$\sim \eta~\la 10^{-9}$), 
the values of $\xi_{i}$ ($i = e, \mu, \tau$) considered here are 
sufficiently small ($|\xi_{i}| \la 0.1$) so that the ``extra'' 
energy density contributed by such degenerate neutrinos is negligible.
\be
\Delta {\rm N}_{\nu}(\xi_{i}) = \frac{30}{7}(\frac{\xi_{i}}{\pi})^{2} 
+ \frac{15}{7}(\frac{\xi_{i}}{\pi})^{4}~\la 0.01.
\ee
In this case, the results to be presented below for $\xi \neq 0$ 
will correspond to \Nnu = 3 ($S = 1$).  In fact, if the three active 
neutrinos ($\nu_{e}$, $\nu_{\mu}$, $\nu_{\tau}$) mix only with each 
other, all individual neutrino degeneracies will equilibrate via 
neutrino oscillations to, approximately, the electron neutrino 
degeneracy before BBN begins~\cite{equilibrate}.  Thus, the magnitude 
of the {\bf electron} neutrino degeneracy constrained by BBN is 
of special interest when limiting the total net lepton asymmetry 
of the universe: $L \approx 3L_{e}$; $\Delta$N$_{\nu}(\xi) \approx 
3\Delta$N$_{\nu}(\xi_{e})$.

For the standard models of particle physics and cosmology, \Deln = 0 
($S = 1$) and $\xi \equiv \xi_{e} \approx \xi_{\mu} \approx \xi_{\tau}
= 0$, and the value (range of values) of $\eta$ identified by BBN and 
the CBR should agree, restricting the allowed deviations from zero of 
\Deln and/or $\xi_{e}$.

\section*{3. CBR}
\begin{figure} [!hbp]
\begin{center}
\includegraphics[clip=true,scale=1.0]{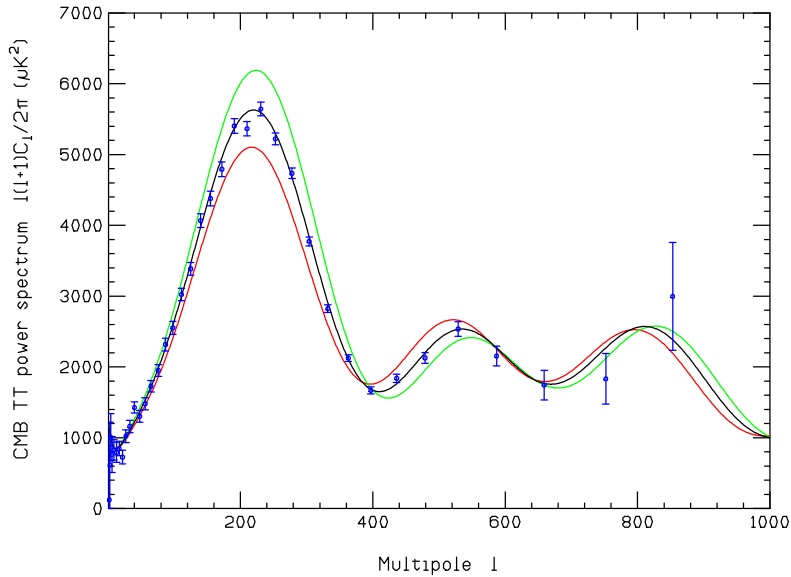}
\end{center}
\caption{The CBR temperature anisotropy angular power spectrum
for three choices of the baryon density, \obh = 0.018, 0.023 
(best fit), 0.028, from bottom to top respectively near $l 
\approx 200$.  The data points with error bars are from Spergel 
{\it et al.}
\cite{sperg}.}
\label{etaspectrum}
\end{figure}
In Figures 1 and 2 are shown the CBR temperature anisotropy angular
power spectra for different choices of the baryon density (Fig.~1)
and of \Nnu (Fig.~2).  Non-zero values of \Deln change the energy
density in radiation, which shifts the redshift of the epoch of equal
matter (Cold Dark Matter and Baryons) and radiation densities.
This results in changes to the angular scales and the amplitudes of 
the ``acoustic'' peaks in Figures 1 \& 2; see, \eg \cite{barger1} 
and further references therein.
WMAP is a much more sensitive baryometer than it is a chronometer.
While the best fit values for the baryon density and \Nnu are
$\eta_{10} = 6.3$ and \Nnu = 2.75 ($S = 0.98$) respectively, 
the $2\sigma$ range for the baryon density parameter is~$5.6 \leq 
\eta_{10} \leq 7.3$, whereas for \Nnu it is~$0.9 \leq {\rm N}
_{\nu} \leq 8.3$ ($0.81 \leq S \leq 1.36$)~\cite{barger2}.
Thus, although the WMAP best fit value of \Nnu is less than the
standard model value of 3, it is clear that this difference is
not at all statistically significant.  It will be interesting 
to see if the much tighter CBR constraint on the baryon density 
parameter ($\sim 6 - 8~\%$) is consistent with the value of this
parameter identified by SBBN.
\begin{figure} [!hbp]
\begin{center}
\includegraphics[clip=true,scale=1.0]{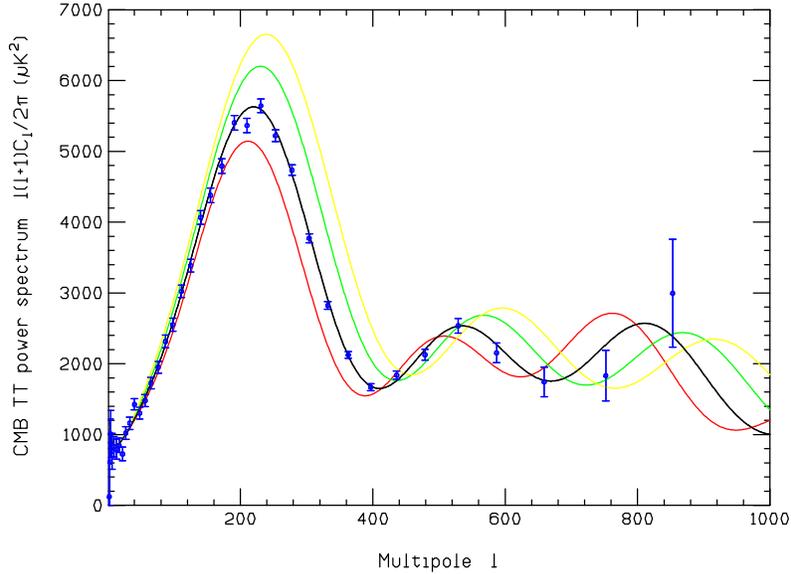}
\end{center}
\caption{The CBR temperature anisotropy angular power spectrum as 
in Fig.~\ref{etaspectrum}, now for four choices of the equivalent 
number of neutrinos \Nnu = 1, 2.75 (best fit), 5, 7, from bottom 
to top respectively near $l \approx 200$.  The data points with 
error bars are from Spergel {\it et al.}
\cite{sperg}.}
\label{nnuspectrum}
\end{figure}

\begin{figure} [!hbp]
\begin{center}
\includegraphics[clip=true,scale=0.42]{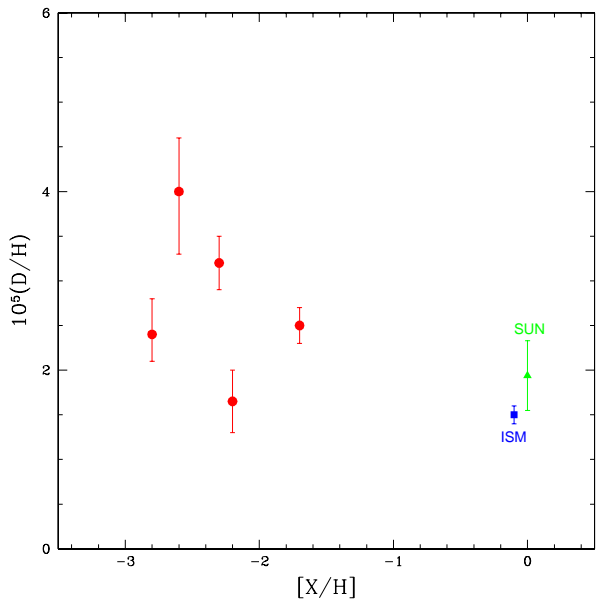}
\end{center}
\caption{The deuterium abundance (by number relative to hydrogen), 
$y_{\rm D} \equiv 10^{5}$(D/H), derived from high redshift, low
metallicity QSOALS~\cite{kirk}~(filled circles).  The metallicity 
is on a log scale relative to solar; depending on the line-of-sight, 
$X$ may be oxygen or silicon.  Also shown is the solar system 
abundance (filled triangle) and that from observations of the 
local ISM (filled square).}
\label{d}
\end{figure}

\section*{4. BBN}

Since the relic abundances of D, \3he, and \7li produced during
BBN are {\it rate} limited (nuclear reaction rates), each of these
nuclides is a candidate baryometer.  Among these, Deuterium is 
the baryometer of choice for several reasons.  The BBN-predicted 
abundance of D is sensitive to the baryon density ($y_{\rm D} \equiv 
10^{5}$(D/H)$_{\rm P} \propto \eta^{-1.6}$).  The post-BBN evolution 
of D is simple in that whenever gas is incorporated into stars, D is 
completely destroyed.  Thus, observations of the deuterium abundance 
anywhere, at any time, provide a {\it lower} bound to the primordial 
D abundance (and, therefore, an {\it upper} bound to $\eta$).  It is 
expected that if D can be observed in regions which have experienced
minimal stellar processing, the deuterium abundance inferred from
such data should be very close to the BBN abundance.  

The good news is that there are data from observations of neutral D 
and neutral H in high redshift, low heavy element (``metallicity'') 
abundance  QSO absorption line systems (QSOALS); see Figure~\ref{d}.  
As may be seen from Fig.~\ref{d}, the bad news is that there are 
only five such systems with good enough data to derive meaningful D 
abundances~\cite{kirk}.  And, even for these, specially selected targets, 
there is the possibility of confusion between D and H absorption spectra 
which are identical, save for the wavelength/velocity shift between them.  
That is, small amounts of hydrogen at the ``wrong'' redshift'' (interlopers) 
can masquerade as deuterium.  Further, since the hydrogen absorption in 
such systems is saturated, it is often difficult to identify the number 
of systems which contribute to the total absorption and this can lead 
to errors in the inferred amount of H in determining the D/H ratio.  
With these caveats in mind, it is important to understand that systematic, 
rather than statistical uncertainties may dominate the error budget.  
Indeed, for the data summarized by Kirkman {\it et al.}~\cite{kirk} 
and shown in Fig.~\ref{d}, $\chi^{2}$ exceeds 16 for 4 degrees of 
freedom!  Following Kirkman {\it et al.}, the error bars are inflated 
here to account for this excessive dispersion, and a primordial 
abundance $y_{\rm D} = 2.6\pm0.4$ is adopted.
\begin{figure} [!hbp]
\begin{center}
\includegraphics[clip=true,scale=0.42]{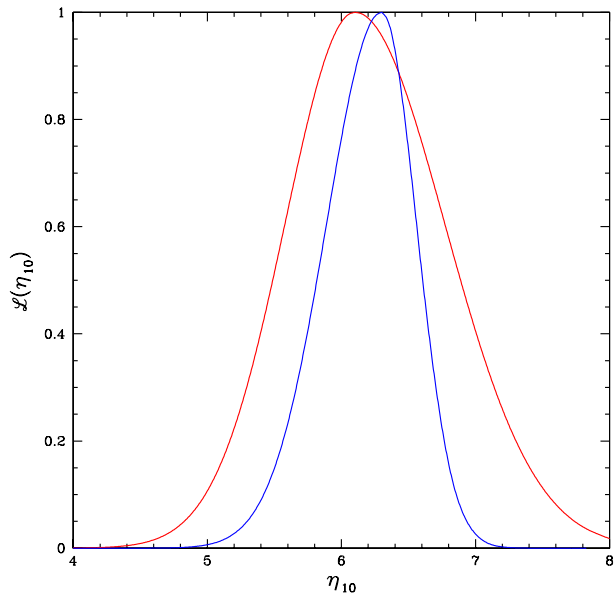}
\end{center}
\caption{The likelihood distributions for the baryon density
parameter $\eta_{10}$ inferred from the CBR (WMAP) and from
SBBN and deuterium; see the text.  The broader distribution,
centered at $\eta_{10} \approx 6.1$ is from SBBN and the 
narrower distribution, centered at $\eta_{10} \approx 6.3$ 
is from the CBR.}
\label{etalikelihood}
\end{figure}
For SBBN this estimate for the abundance of primordial D corresponds 
to $\eta_{10} =  6.1^{+0.7}_{-0.5}$.  This is in excellent (spectacular!)
agreement with the completely independent estimate above in \S3 from
the CBR.  In Figure~\ref{etalikelihood} are shown the likelihood
distributions for $\eta$ derived from the CBR (WMAP) and from SBBN 
(D).

\3he is a much less useful baryometer than is D.  In the first place,
its BBN-predicted abundance is less sensitive to the baryon density
parameter (\3he/H $\propto \eta^{-0.6}$).  In addition, as gas is 
incorporated into stars and the stellar-processed material is returned 
to the interstellar medium when they die, \3he is produced, destroyed 
and, some survives.  This complicated history makes it much more difficult 
to account for the post-BBN evolution of \3he.  Finally, \3he is only 
observed within the Galaxy~\cite{brb} and, therefore, its abundance 
samples only a limited range in metallicity.  Nonetheless, while there 
is a clear oxygen abundance gradient with location within the Galaxy 
(higher in the center, lower in the suburbs, indicating more stellar 
processing in the interior), the \3he abundance shows no such gradient 
(either with position or with metallicity).  However, the Bania, Rood,
and Balser~\cite{brb} recommended value of $y_{3} \equiv 10^{5}$(\3he/H) 
$= 1.1\pm0.2$ corresponds to $\eta_{10} \approx 5.6^{+2.0}_{-1.7}$,
in broad agreement with the SBBN-D and CBR-WMAP determinations.  So
far, so good.

In contrast to deuterium (and to \3he), the abundance of \4he has 
increased in the post-BBN universe as stars have burned hydrogen to 
helium (and beyond).  Therefore, to avoid model-dependent evolutionary 
uncertainties, it is best to concentrate on determining the \4he
abundance (mass fraction, Y$_{\rm P}$) in the most metal-poor sample
available and to let the data speak for themselves concerning any 
correlation between the helium and heavy element abundances.  The 
best such data come from observing the emission lines formed when 
ionized helium and hydrogen capture electrons in regions of hot, 
ionized gas (\hii regions).  There exists at present, thanks largely 
to the work of Izotov and Thuan~\cite{it}, a very large sample of 
helium abundance determinations in low metallicity, extragalactic 
\hii regions.  This newer, more uniform data set complements earlier, 
more heterogeneous samples~\cite{he4}.  For the WMAP estimate of 
the baryon density, including its uncertainty, the SBBN-predicted 
primordial abundance of \4he is \Yp $= 0.2482\pm0.0007$.  Unfortunately, 
{\it none} of the \Yp estimates~\cite{it,he4} agree with the SBBN 
prediction, all being low by roughly 2$\sigma$.  Indeed, from their 
2004 sample of 82 data points Izotov \& Thuan~\cite{it} derive such 
a small uncertainty, that their central value is low by nearly 
6$\sigma$!  For the numerical results presented in \S5 below, a 
primordial mass fraction \Yp = 0.238$\pm0.005$~\cite{osw} is adopted.

The \4he abundance determinations are most likely examples of 
extremely precise, yet inaccurate, determinations of an important
cosmological parameter.  It has long been known that there are a
variety of systematic uncertainties which are likely to interfere
with an accurate \Yp determination.  In a very recent, detailed 
study of {\it some} of these identified systematic uncertainties, 
Olive \& Skillman~\cite{os} suggest the true errors likely exceed
previous estimates by factors of 2--3 or more ($\Delta$Y$_{\rm P} 
\approx 0.013$).  Given such large uncertainties, it is not surprising
that the extant \4he abundance data are, within the errors, consistent
with the predictions of SBBN and the CBR (and/or SBBN plus D) determined
baryon density.  Nonetheless, it might be premature to ignore this
challenge to the standard models of particle physics and of cosmology.
Perhaps the tension between D and \4he is an early warning of new 
physics beyond the standard models.  Before pursuing this option in
the next section, \7li is considered here.

As with \4he, there is conflict in the comparisons between the SBBN
predictions and the \7li relic abundance estimates derived from 
observations.  Here, too, the potential for systematic errors looms 
large.  \7li is produced in the Galaxy by cosmic ray spallation/fusion
reactions and observations of super-lithium rich red giants provide 
evidence that (at least some) stars can be net producers of lithium.  
Therefore, to infer the BBN yield of \7li, the data should  be 
limited to those from the most metal-poor halo stars in the Galaxy.  
For the WMAP baryon density, the SBBN expected \7li abundance is 
[Li]$_{\rm P} \equiv 12 + $log(Li/H)$_{\rm P} = 2.65^{+0.05}_{-0.06}$ .
In contrast, for a selected data set of the lowest metallicity halo 
stars, Ryan \etal \cite{ryan} derive a primordial abundance of 
[Li]$_{\rm P} \approx 2.0-2.1$.  In deriving the stellar lithium 
abundances, the adopted stellar temperature plays a key role.  When 
using the infrared flux method effective temperatures, studies of 
halo and Galactic Globular Cluster stars \cite{bonif} suggest a 
higher abundance: [Li]$_{\rm P} = 2.24\pm0.01$.  Very recently, 
Melendez \& Ramirez~\cite{mr} have reanalyzed 62 halo stars using 
an improved infrared flux method effective temperature scale, confirming 
the higher lithium abundance; they find [Li]$_{\rm P} = 2.37\pm0.05$.  
If this were the true primordial \7li abundance, then the SBBN value
of the baryon density parameter would be $\eta_{10} = 4.5\pm0.4$, 
in conflict with the CBR-WMAP and/or SBBN-D estimates.  Indeed, 
all of the current observational estimates of the abundance of
primordial lithium are significantly lower than the SBBN expectation.

As with \4he, the problem may be traced to the astrophysics rather 
than to the cosmology.  Since the low metallicity halo stars used 
to estimate the primordial abundance of lithium are the oldest stars 
in the Galaxy, they have had the most time to modify (by dilution 
and/or destruction) their surface abundances.  While mixing of the 
stellar surface material with the interior would destroy or dilute 
prestellar lithium, the very small dispersion among the observed 
values of [Li] derived from the lowest metallicity halo stars suggests 
this effect may not be large enough to bridge the $\approx 0.3$~dex 
gap between the observed and CBR/SBBN-predicted abundances; see, 
\eg \cite{pinsono} and further references therein.
 
\section*{5. CBR And BBN Combined}

\begin{figure} [!hbp]
\begin{center}
\includegraphics[clip=true,scale=0.65]{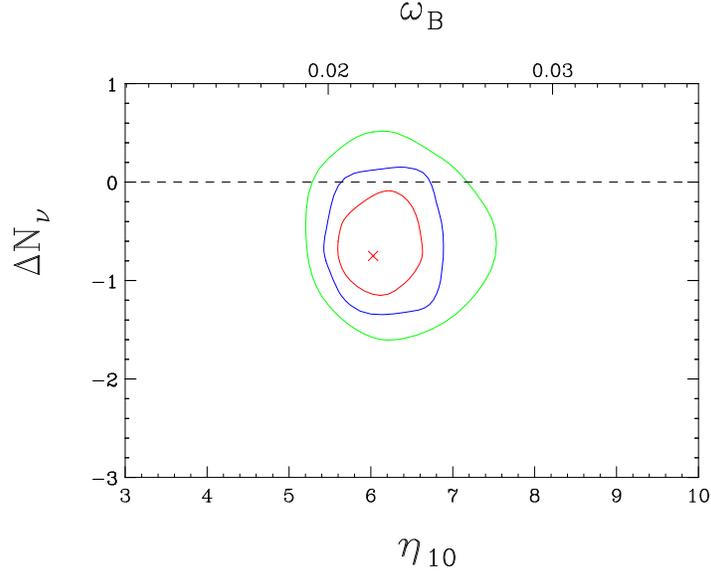}
\end{center}
\caption{The $1\sigma$, $2\sigma$, and $3\sigma$ contours in the
\Deln -- $\eta_{10}$ plane (from Barger {\it et al.}~\cite{barger1}) 
consistent with the WMAP CBR data and the BBN predicted and observed 
abundances of D and \4he.  The cross marks the best fit point; see 
the text.  Note that the upper horizontal axis is for $\omega_{\rm B} 
\equiv \Omega_{\rm B}h^{2}$.}
\label{jointbbncmb}
\end{figure}

In contrast to D (and \3he and \7li), \4he is an insensitive baryometer,
but its primordial abundance is a useful, early universe chronometer.
If the standard model expansion rate is changed ($S \neq 1$, \Deln $\neq
0$), this will affect the neutron-proton ratio at BBN and change the
SBBN-expected value of Y$_{\rm P}$.  The current conflict between the
SBBN-predicted and the observationally inferred values of \Yp requires
a slowdown in the early universe expansion rate ($S < 1$; \Deln $< 0$).
A joint BBN fit to the observationally inferred D and \4he abundances
suggests that $\eta_{10} \approx 5.7$ and $S \approx 0.94$ (\Deln 
$\approx -0.70$) can relieve the SBBN tension between D and \4he.  
However, it can be seen in Figure~\ref{jointbbncmb} that while 
these values are entirely consistent with the constraints from 
the CBR, \Nnu = 3 is consistent with both BBN and the CBR at 
$\sim 2\sigma$~\cite{barger1}.  Nonetheless, this combination of 
parameters does not resolve the conflict with \7li.  Although a 
slowdown in the expansion rate has the effect of increasing \7li 
(more time for production), this is compensated by the somewhat 
lower baryon density (slower reaction rates), which has the opposite 
effect.  The result is that for the choices of $S$ and $\eta$ which 
resolve the conflicts between D and \4he (and between WMAP and \4he), 
the predicted primordial abundance of \7li is [Li]$_{\rm P} \approx 
2.62\pm0.10$, still some 0.2--0.3 dex higher than that inferred 
from the data. 

Another example of new physics with the potential to resolve the 
D --\4he conflict while leaving the CBR--D agreement unaffected is 
a neutrino-antineutrino asymmetry ($\xi \equiv \xi_{e} \neq 0$);
see~\cite{barger2} and references therein.
\begin{figure} [!hbp]
\begin{center}
\includegraphics[clip=true,scale=0.55]{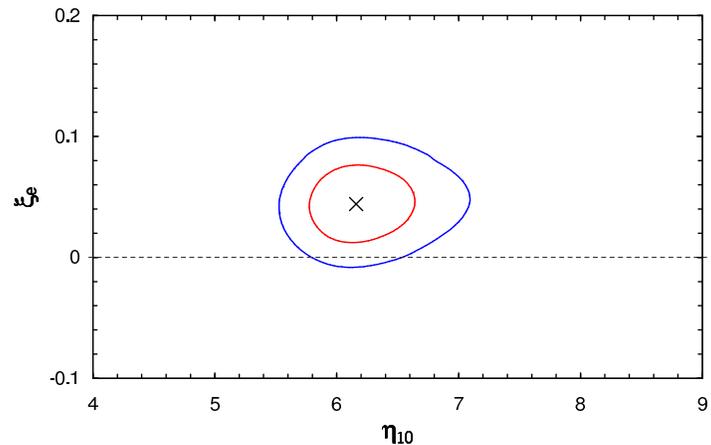}
\end{center}
\caption{The $1\sigma$ and $2\sigma$ contours in the
$\xi_{e} - \eta_{10}$ plane (from Barger {\it et al.}~\cite{barger2}) 
consistent with the WMAP CBR data and the BBN predicted and observed 
abundances of D and \4he.  The cross marks the best fit point; see 
the text.}
\label{xivsetannu}
\end{figure}
Through its effect on the neutron-proton ratio at BBN, such an asymmetry
can change the BBN-predicted \4he abundance.  For $\xi_{e} > 0$, there
are more $\nu_{e}$ than $\bar{\nu}_{e}$ and this drives the n--p ratio
down, decreasing the BBN-predicted \4he abundance.  For $\xi_{e}~\la 
0.1$, the extra energy density contributed by these degenerate neutrinos
is small, and \Nnu = 3 remains a good approximation.  As a result, such
a lepton asymmetry has negligible effect on the CBR temperature anisotropies
and the good agreement with the WMAP data is unaffected.  The effects 
of such an asymmetry on the predicted abundances of D, \3he, and \7li 
are subdominant to that on \4he.  The best fit parameter choices 
which resolve the tension between D and \4he, while preserving the good 
agreement with the WMAP data are $\eta_{10} = 6.2$ ($\Omega_{\rm B}h^{2} 
= 0.23$) and $\xi_{e} = 0.044$.  In Figure~\ref{xivsetannu} are shown 
the $1\sigma$ and $2\sigma$ contours consistent with the WMAP data and 
with BBN (D and \4he).

Finally, we note that if all three parameters ($\eta$, $S$, and $\xi$)
are allowed to be free, much larger ranges in them will remain consistent
with BBN (D and \4he) and with the CBR; see, \eg \cite{barger2} and the
more recent paper of Kneller and Steigman~\cite{ks}.  Fixing the D and
\4he abundances, Kneller \& Steigman find two approximate, but quite
accurate, BBN relations among these three parameters.
\be
590(S - 1) \approx 116\eta_{10} - 697,
\ee
and
\be
145\xi_{e} \approx 106(S - 1) + 6.31.
\ee
Consistent with the WMAP CBR data, values of \Deln and $\xi_{e}$ in the 
ranges $-2~\la \Delta$N$_{\nu}~\la +5$ and $-0.1~\la \xi{_e}~\la +0.3$
are permitted~\cite{barger2}.  However, even with this freedom it is
still not possible to reconcile the BBN-predicted and the observed
relic lithium abundances~\cite{ks}.  For the values of these parameters
which are consistent with BBN and the CBR, [Li]$_{\rm P} \approx 2.6$.  

\section*{6. Summary and Conclusions}

BBN and the CBR probe the evolution of the Universe (and its constituents)
at two, widely separated epochs in its early evolution.  Confronting the 
predictions of BBN and the CBR with the relic abundance and WMAP data 
enables independent tests of the standard models of particle physics 
and cosmology.  Qualitatively, the standard models pass these tests 
with flying colors, permitting BBN and CBR constraints to be put on
new neutrinos physics (N$_{\nu} \neq 3$?; $\xi_{e} \neq 0$?).  When 
considered in quantitative detail however, there are some challenges 
to the standard models at the $\sim 2\sigma$ level.  Many would take 
this as evidence for success and declare victory.  However, if these 
tensions are taken at face value, they might be alerting us to problems 
with the astronomy, the astrophysics, the cosmology, the particle physics 
or, combinations of them.  It is not unlikely that the apparent conflicts 
may result from the data, its analysis, and/or the extrapolations to 
the early universe.  While the community awaits the new surprises to 
be encountered at the LHC, a Linear Collider, or the next generation 
of terrestrial or space-based telescopes, it should be kept in mind 
that these challenges could be pointing the way to new physics, 
especially new neutrino physics,  beyond the standard models of 
particle physics and/or cosmology.

\newpage

\bigskip\noindent
{\large \bf Acknowledgements}

I wish to express my sincere appreciation and thanks to the organizers
of this symposium and to the Nobel Foundation for its sponsorship.
Special thanks are due Per Olof Hulth and Tommy Ohlsson for their
tireless efforts to smooth the way and ensure that my participation
would be so enjoyable and scientifically successful.  The research 
described here has been supported at The Ohio State University by 
a grant from the US Department of Energy (DE-FG02-91ER40690).

\end{document}